\documentclass[journal=ancac3,manuscript=article,layout=twocolumn]{achemso}

\usepackage{braket}
\usepackage{amssymb}
\usepackage{amsmath}
\usepackage{textcomp}
\usepackage[normalem]{ulem}
\usepackage{chemformula} 
\usepackage[T1]{fontenc} 

\author{Sergey A. Dyakov}
\email{s.dyakov@skoltech.ru}
\affiliation{Skolkovo Institute of Science and Technology, Nobel Street 3, Moscow, Russia}
\author{Margarita V. Stepikhova }
\affiliation{Institute for Physics of Microstructures RAS, GSP-105, Nizhny Novgorod, 603950, Russia}
\author{Andrey A. Bogdanov}
\affiliation{Department of Physics and Engineering, ITMO University, Birjevaja line V.O. 14, St. Petersburg, 199034, Russia}
\author{Alexey V. Novikov}
\affiliation{Institute for Physics of Microstructures RAS, GSP-105, Nizhny Novgorod, 603950, Russia}
\altaffiliation{Lobachevsky State University of Nizhny Novgorod, Gagarin ave. 23, Nizhny Novgorod, 603950, Russia}
\author{Dmitry V. Yurasov}
\affiliation{Institute for Physics of Microstructures RAS, GSP-105, Nizhny Novgorod, 603950, Russia}
\author{Zakhary F. Krasilnik}
\affiliation{Institute for Physics of Microstructures RAS, GSP-105, Nizhny Novgorod, 603950, Russia}
\altaffiliation{Lobachevsky State University of Nizhny Novgorod, Gagarin ave. 23, Nizhny Novgorod, 603950, Russia}
\author{Sergei G. Tikhodeev}
\affiliation{Lomonosov Moscow State University, Leninskie Gory, GSP-1, Moscow, 119991, Russia}
\altaffiliation{A.M. Prokhorov General Physics Institute, Vavilova st. 38, Moscow, 117942, Russia}
\author{Nikolay A. Gippius}
\affiliation{Skolkovo Institute of Science and Technology, Nobel Street 3, Moscow, Russia}

\title{Photonic bound states in the continuum in Si structures with the self-assembled Ge nanoislands}
\abbreviations{
BIC --- bound state in the continuum\\
DPL --- directional photoluminescence\\
FMM --- Fourier modal method\\
FW BIC --- Friedrich-Wintgen bound state in the continuum\\
$\mu$PL --- microphotoluminescence\\
PCS --- photonic crystal slab\\
PL --- photoluminescence\\
SP BIC --- symmetry protected bound state in the continuum\\
TE --- transverse electric\\
TM --- transverse magnetic}


\begin{document}

\begin{tocentry}

\includegraphics{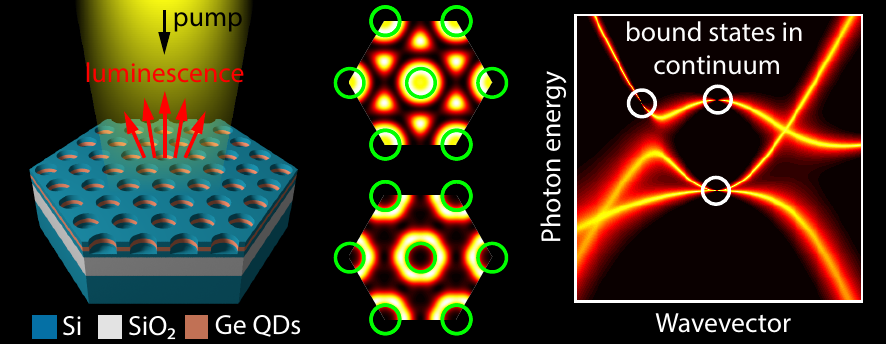}

\end{tocentry}

\begin{abstract}
Germanium self-assembled nanoislands and quantum dots are very prospective for CMOS-compatible optoelectronic integrated circuits but their luminescence intensity is still insufficient for many practical applications. Here, we demonstrate experimentally that photoluminescence of Ge nanoislands in silicon photonic crystal slab with hexagonal lattice can be dramatically enhanced due to the involvement in the emission process of the bounds states in the continuum. We experimentally demonstrate more than two orders of magnitude peak photoluminescence enhancement and more than one order of magnitude integrated PL enhancement in a photonic crystal slab compared with the non-structured sample area. We theoretically study this effect by the Fourier modal method in the scattering matrix form and demonstrate the appearance of quasi-normal guided modes in our photonic crystal slab. We also describe their symmetry in terms of group theory. Our work paves the way towards a new class of optoelectronic components compatible with silicon technology.
\end{abstract}

\section{Introduction}

Silicon technology is the base of modern nanoelectronics. In spite of active search of an alternative platform like plasmonics, polaritonics, graphene electronics, etc., it is hard to imagine that some of them can replace silicon technology in the nearest future. One of the main challenges of {\em silicon photonics} is a lack of effective light sources that can be incorporated in CMOS-compatible integrated circuits. A plethora of possible candidates for active media in silicon photonics was analyzed in detail for the last four decades. 

In particular, much attention was paid to Er-doped silicon structures giving luminescence peak around 1.55~$\mu$m~\cite{Ennen1983, Ennen1985}. However, due to a rather long spontaneous transition lifetime and the limited solid solubility of Er in Si, photoluminescence efficiency is quite low~\cite{Kenyon2005,Xie1991, Huda2003}. Another prospective active medium compatible with silicon technology is n-Ge strained layers giving  luminescence in the telecommunication spectral range. Application of tensile strain and/or heavy $n$-type doping can effectively reduce the energy difference between the direct and indirect transitions in Ge increasing the probability of radiative recombination \cite{Liu2007}. Using heavy doping and low strain lasing in Ge was demonstrated both under optical~ \cite{liu2010ge} and electrical~ \cite{camacho2012electrically} pumping. However, the thresholds were unpractically large. Subsequent application of higher strains allowed to achieve laser action at much smaller thresholds but at cryogenic temperatures~\cite{bao2017low, elbaz2018germanium, pilon2019lasing}.

Direct-bandgap transitions can be achieved in GeSn alloys offering a tunable bandstructure~\cite{sun2012towards, chen2014demonstration, wirths2015lasing, stange2016optically, chretien2019gesn, Elbaz2020}. Luminescence can be also obtained from Si itself due to the quantum confinement spreading out the carrier wavefunction in the momentum space increasing the probability of radiative processes. Such a luminescence was demonstrated in porous Si~\cite{Koshida1992, Hirschman1996}, Si nanocrystals~\cite{Cullis1991, Wilson1993, Pavesi2000, valenta2019nearly, PhysRevB.93.205413}, and Si/Ge quantum wells~\cite{Dehlinger2000}. Light-emitting A$_3$B$_5$ structures can be integrated to silicon platform via direct epitaxial growth~\cite{bolkhovityanov2008gaas, li2017epitaxial, liu2018photonic, norman2019review} or wafer bonding techniques~\cite{fang2007hybrid, roelkens2007iii, park2011device}. In spite of intensive study, none of the named above luminescence mechanisms found large-scale implementation in industry. 



\begin{figure}[b!]
\centering
\includegraphics[width=1\linewidth]{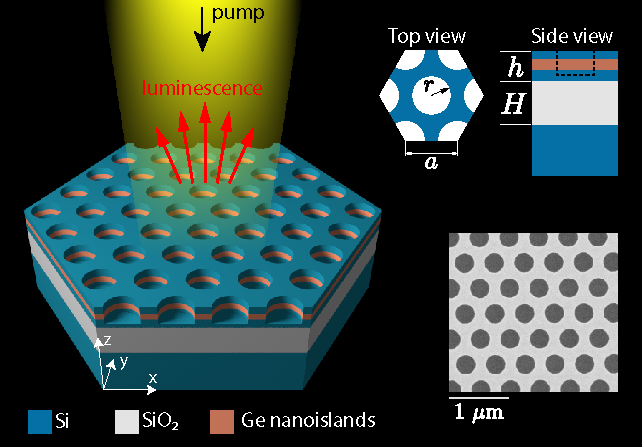}
\caption{Schematics of the PCS with Ge nanoislands. Insets show the top and side views of the PCS and the SEM image of the PCS with air pores in Si matrix.}
\label{intro}
\end{figure}


An alternative CMOS-compatible active medium is Ge self-assembled nanoislands. Room-temperature photoluminescence (PL) is observed in such structures at wavelengths 1.3--1.6\,$\mu$m (0.75--0.95\,eV). However, due to the spatial separation of holes and electrons in Ge nanoislands \cite{aleshkin1998self}, their radiative recombination efficiency is not high enough for practical applications. There are several approaches for the increase of this efficiency among which may be mentioned a vertical arrangement of nanoislands in a lattice~\cite{Talalaev2006} and ion bombarding of Ge nanoislands~\cite{Grydlik2016}. Thus, the low-temperature laser generation was demonstrated in whispering gallery mode resonators with built-in Ge nanoislands~\cite{Grydlik2016}.

Photoluminescence from Ge nanoislands can be  enhanced via the Purcell effect in various resonant photonic structures including Mie resonators~\cite{Rutckaia2017}, photonic crystal cavities~\cite{Sylvain2004, Stepikhova2019}, metasurfaces~\cite{Zeng2016, Yuan2017}, nanoantennas~\cite{Zeng2016}, etc. An important advantage of Ge nanoislands is that they can be precisely positioned at the hotspots of the mode in photonic structures~\cite{schatzl2017enhanced}. Photonic crystal cavities demonstrate extremely high $Q/V$ values and, thus, high Purcell factor  offering a lot of benefits for compact optical devices with strong light-matter interaction~\cite{Quan2011highQ,Yoshie2004}. From the other side, due to a small active region of photonic crystal cavities in comparison with the overall footprint, they are not very promising for light-emitting applications.    

Periodic photonic structures without cavities can support high-Q states with a mode profile homogeneously spread over the whole photonic structure. Such states are now called {\it bound states in the continuum} (BICs) and recently they become to attract enormous attention in photonics~\cite{Hsu2016a, Zhen2014, rybin2017optical, koshelev2019light, jin2019topologically, koshelev2019meta}. BICs represent spatially localized states with vanishing radiation despite their energy embedded in the continuum spectrum of the environment. Fundamentally, BICs originate from destructive interference, when two or more waves superpose to completely suppress radiative losses~\cite{koshelev2020engineering, koshelev2019nonradiating, yang2014analytical, sadrieva2019multipolar}. Therefore, their radiation lifetime diverges in theory. However, in practice, due to the finite size of the sample, roughnesses, and other imperfections, the radiative Q factor of BIC becomes finite but extremely large~\cite{Hsu2013, bulgakov2019high, bulgakov2017light}.  BICs were first predicted in quantum mechanics around a century ago\cite{vonNeumann1929} but in optics, they have been actively studied over the last decade~\cite{Pacradouni2000, Paddon2000, Marinica2008, Bulgakov2008}. The close attention paid to BICs is explained by a variety of their potential applications for resonant field enhancement~\cite{Mocella2015, Magnusson2015}, lasing~\cite{Kante2017laser, Bahari2018, ha2018directional}, filtering of light~\cite{Foley2014, Cui2016scirep}, biosensing~\cite{romano2018label, romano2018optical, Liu2017bic}, enhancement of light-matter interaction~\cite{kravtsov2020nonlinear, koshelev2018strong}, polarization control~\cite{doeleman2018experimental, zhang2018observation}, and non-linear photonics~\cite{ koshelev2019nonlinear, krasikov2018nonlinear, bulgakov2019nonlinear, bulgakov2011symmetry, bulgakov2013channel}. The mechanism resulting in the appearance of BIC in periodic structures can be exploited to engineer high-Q states (quasi-BIC) in single resonators~\cite{rybin2017high, koshelev2020subwavelength, mylnikov2020lasing, bogdanov2019bound}.

 
Here, we use the BICs to enhance photoluminescence (PL) from Ge nanoislands grown by molecular beam epitaxy on silicon-on-insulator wafer and embedded in photonic crystal slab (PCS) (Figure\,\ref{intro}). We experimentally demonstrate more than two orders of magnitude peak photoluminescence enhancement and more than one order of magnitude integrated PL enhancement in a photonic crystal slab compared with the non-structured sample area. 

\section{PL enhancement and quasiguided modes}

\begin{figure}[b!]
\centering
\includegraphics[width=0.85\linewidth]{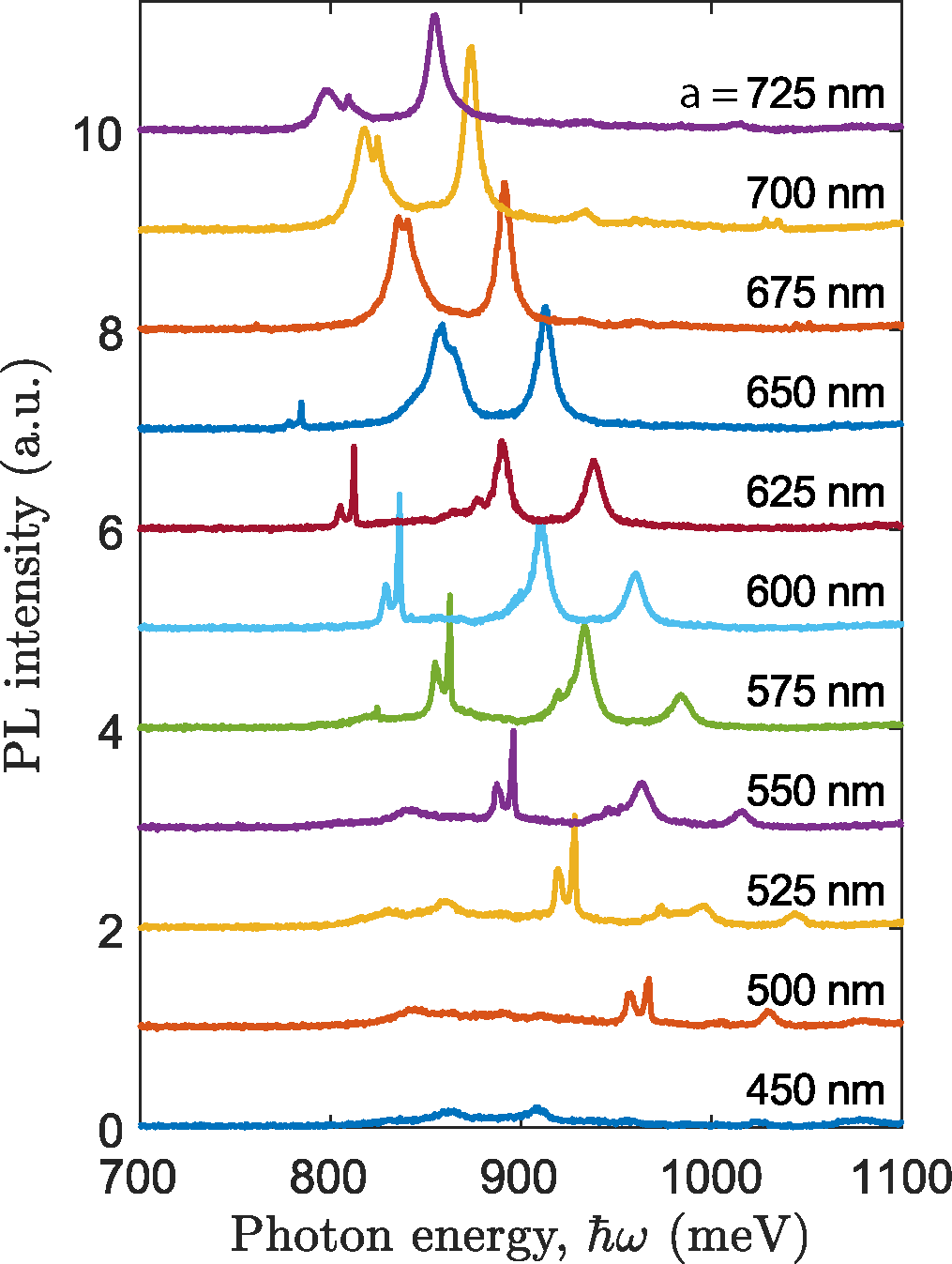}
\caption{Experimental PL spectra for different photonic crystal periods measured at room temperature in DPL scheme. For all spectra $r/a=0.2$.}
\label{spectra3}
\end{figure}

To study the effect of the PL enhancement we use the PCS with hexagonal photonic crystal lattice of air pores (Figure\,\ref{intro}) formed in SOI wafer with the  thickness of buried oxide $H=3$\,$\mu$m. The thickness of the whole structure above the buried oxide was $h = 300$\,nm, which included the 60\,nm-thick Ge nanoislands lattice consisting of 5 layers with Ge nanoislands separated by 15\,nm Si spacer layers. Such lattice was sandwiched between the 75\,nm and 165\,nm-thick capping and buffer Si layers, respectively. The lattice period of a photonic crystal, $a$, was varied in the range from 450 to 725\,nm, and the ratio of the pore radius to period was $r/a=0.2$ and 0.26. We measure the PL spectra of the PCSs using two different schemes, namely, the microphotoluminescence ($\mu$PL) and directional photoluminescence (DPL) schemes. The main difference between them is the solid angle to the surface normal from which the PL signal is detected (see Methods).

\begin{figure*}[t!]
\centering
\includegraphics[width=1\linewidth]{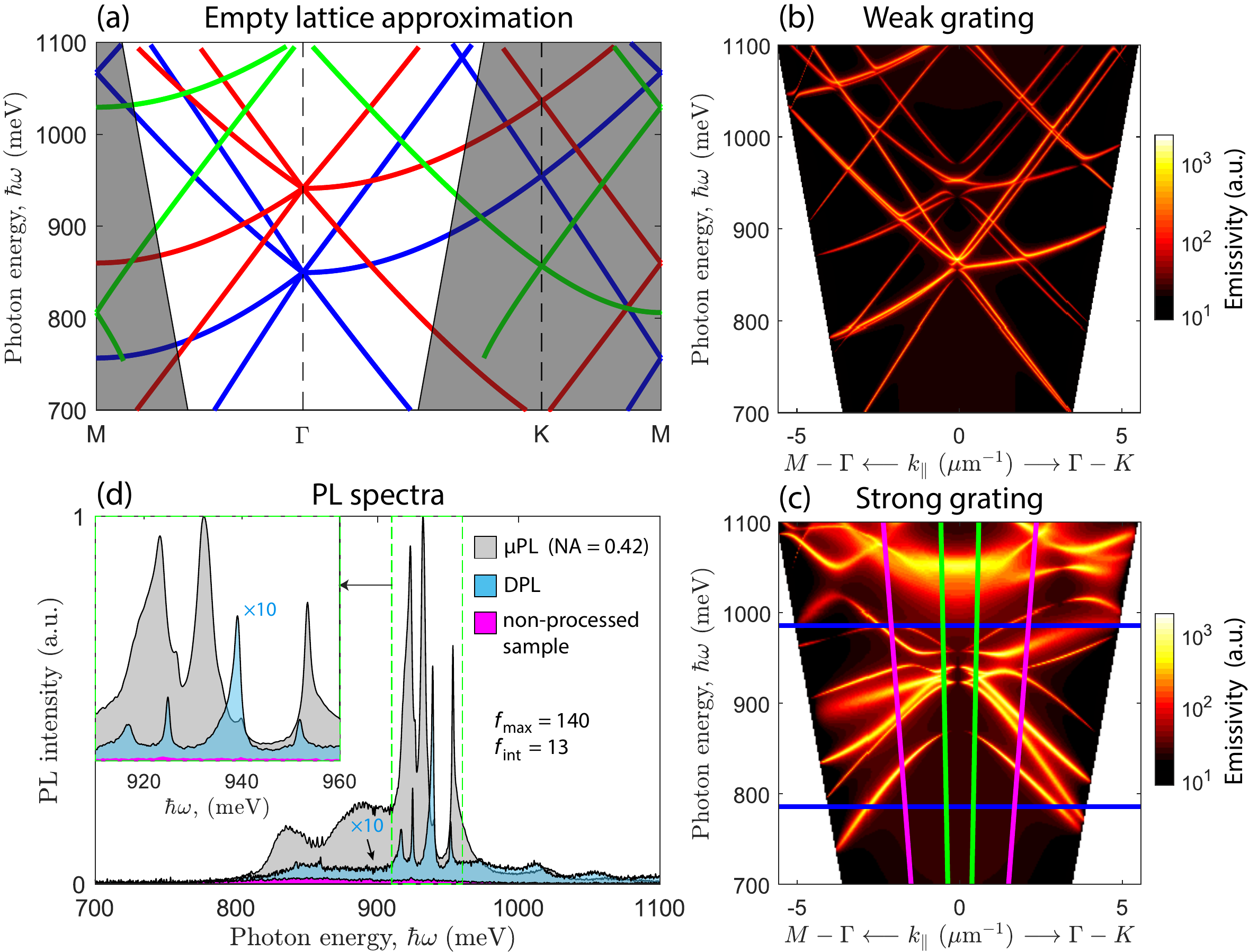}
\caption{(Color online) (a) Dispersion of the quasiguided modes of the PCS in empty lattice approximation with PCS period $a=570$\,nm. Blue, red, and green colors denote, respectively, TE$_1$, TM$_1$, and TE$_2$ quasiguided modes. Gray color denotes the region below the vacuum light line. (b--c) Photon energy and in-plane wavevector dependence of emissivity calculated for weak and strong gratings with $a=570$\,nm and $r/a=0.26$. Green and magenta lines in panel (c) denote the 6$^\circ$ and 25$^\circ$ light cones from which the PL light is collected in DPL and $\mu$PL  techniques, respectively. Blue lines bound the energy range of intrinsic photoluminescence of Ge nanoislands. Maps in panels (b) and (c) are presented in the logarithmic color scale shown on the right. (d) PL spectra measured for the PCS with $a=570$\,nm and $r/a=0.26$ using $\mu$PL and DPL techniques. Inset in (d) shows the same spectra on an enlarged scale. The peak $\mu$PL enhancement is $f_{\mathrm{max}}=140$, integral $\mu$PL enhancement is $f_{\mathrm{int}}=13$.}
\label{EKX}
\end{figure*}
\begin{figure*}[t!]
\centering
\includegraphics[width=1\linewidth]{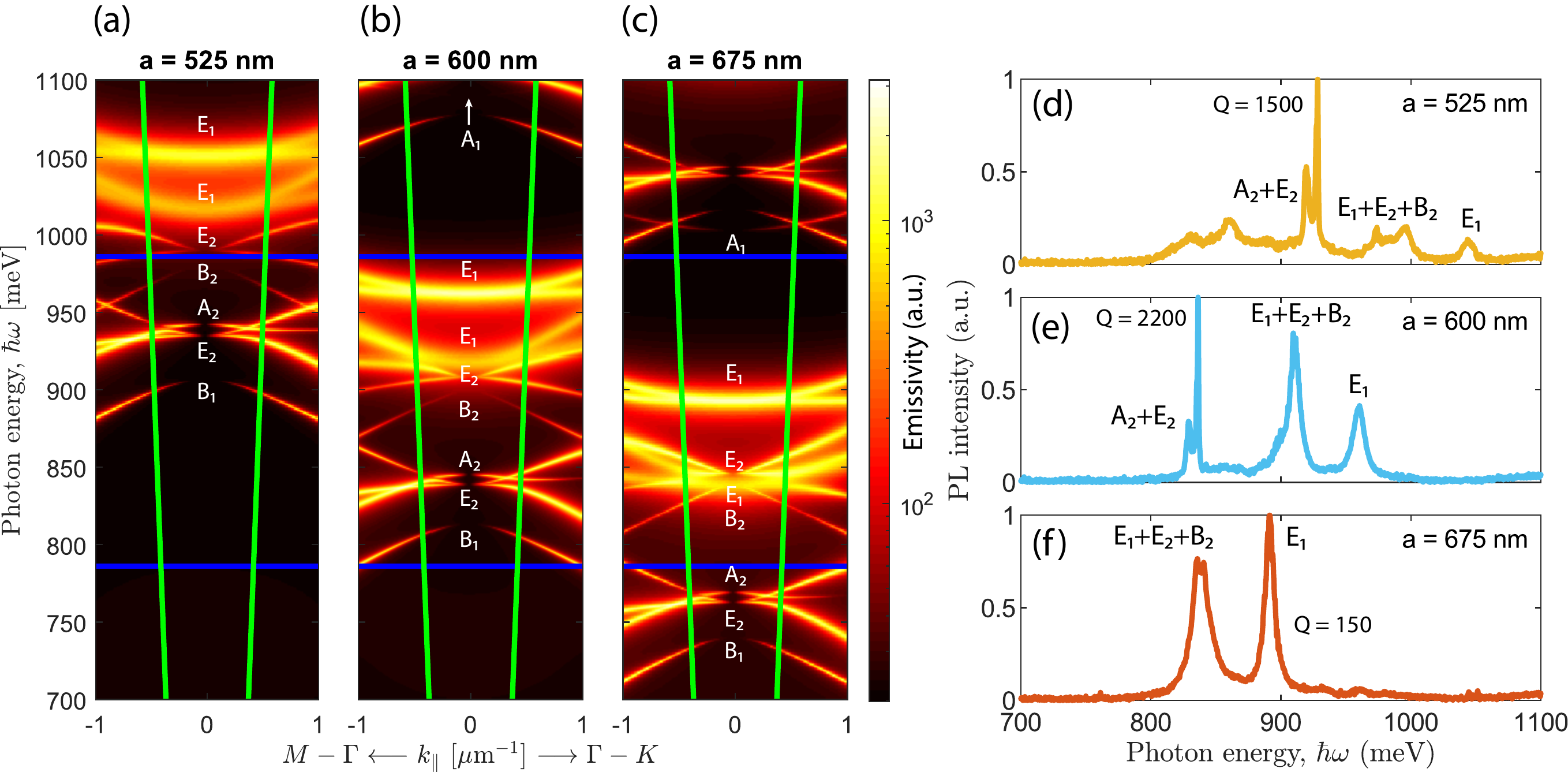}
\caption{(a--c) Calculated emissivity near $\Gamma$-point at $r/a=0.2$ for different photonic crystal periods. The logarithmic color scale shown on the right. Green lines denote the 6$^\circ$ light cone from which the PL light is collected. Horizontal blue lines bound the energy range of intrinsic photoluminescence of Ge nanoislands. (d--f) Experimental PL spectra measured in the DPL scheme for different photonic crystal periods.}
\label{spectra4}
\end{figure*}

Experimental PL spectra measured with DPL technique (collection angle $\pm$6$^\circ$ to the surface normal) at room temperature for PCS with different periods are shown  in Figure\,\ref{spectra3}. One can see that each PL spectrum from PCS consists of several resonance peaks. The experimental peaks have the following features: (i) there are narrow and wide peaks; (ii) some of the peaks have a fine structure; (iii) peaks have a profile which is close to a Laurentian. The peaks spectral position redshifts with an increase of the lattice period. It is well known that photonic crystal slabs have such peaks in their optical spectra of reflection, transmission, and photoluminescence \cite{Tikhodeev2002b, fan2002analysis}. They represent the quasiguided modes \cite{Tikhodeev2002b}, (also known as quasi-normal guided modes \cite{gras2019quasinormal,Lalanne2019}) that appear due to the grating assisted coupling of waveguide modes with photon continuum of the far field. To establish the cause of differences in the peaks, below we study the nature and symmetry of the modes of our hexagonal lattice in detail.  

We start from the empty lattice approximation and plot the resonances of the effective homogeneous waveguide folded into the first Brillouin zone of the hexagonal lattice (Figure\,\ref{EKX}a). In the displayed spectral range, one can observe several families of modes in the $\Gamma$-point. They correspond to the TE$_1$, TM$_1$, and TE$_2$ waveguide modes (blue, red, and green curves, respectively) and the second-order TE mode (green curve). With the introduction of the pores, these modes start to interact with each other which results in splitting and bending of their dispersion curves. This is illustrated in Figure\,\ref{EKX}b, where we present the calculated photon energy, $\hbar\omega$, and in-plane wavevector, $k_\parallel$, the dependence of the emissivity of a hypothetical low-contrast (weak) grating where the dielectric permittivity of the substance in pores is only 20\% less than the dielectric permittivity of the matrix. One can see that in the weak grating the modes in $\Gamma$-point are split, but the dispersions of quasiguided modes can be described roughly in the empty lattice approximation. The number of the first order quasiguided modes in the hexagonal photonic crystal lattice is 12, although not all of them can be distinguished in the dispersion diagram in Figure\,\ref{EKX}b as some of them are still too close to each other. One can see in Figure\,\ref{EKX}b that in the $\Gamma$-point some of the modes are degenerate while out of the $\Gamma$-point the degeneracy is lifted. The group theory predicts for our C$_{6v}$ symmetric hexagonal PCS that in the $\Gamma$-point there are doubly degenerate modes (i.e., doublets) and non-degenerate modes (i.e., singlets). The number of singlets and doublets is determined by a symmetry of the photonic crystal lattice. Here there are four first-order singlets and four first order doublets. With the increase of the grating contrast, the modes continue to hybridize forming an even more complicated dispersion diagram (Figure\,\ref{EKX}c). However, the number of modes and their degree of degeneracy in the $\Gamma$-point remain unchanged.

Since all the resonances have a dispersion with $k_\parallel$, the resonance peaks, which appear in the measured PL spectra, inhomogeneously broaden due to a non-zero numerical aperture (NA) of the collecting lens. One can reduce the effect of such aperture by using the lens with a smaller NA. This can be understood by inspecting Figure\,\ref{EKX}c where the green and magenta lines bound 6$^\circ$ and 25$^\circ$ light cones from which the PL light is collected in our DPL and $\mu$PL setups, respectively (see Methods). The corresponding experimental PL spectra for PSC with $a=570$\,nm and $r/a=0.26$  are shown in Figure\,\ref{EKX}d. One can see from the presented spectra that the position of the resonance peaks coincides quite well with the calculated position of the modes near the $\Gamma$-point in Figure\,\ref{EKX}c. When the PL is measured from inside the 25$^\circ$ cone, the resonance peak positions are averaged over the wide range of emission angles. It leads to a stronger integrated PL signal and broader resonance peaks. The resulting PL intensity of the PCS is enhanced by a factor of $f_{\mathrm{max}}=140$ at $\hbar\omega=932$\,meV (1330\,nm) compared to the PL intensity of the non-processed film. The spectrally integrated enhancement factor is $f_{\mathrm{int}}=13$. Whereas the PL spectra measured from inside the 6$^\circ$ cone have a weaker PL intensity but narrower peaks. Moreover, the peak positions in the DPL spectra correspond more closely to the calculated PCS quasiguided modes in the $\Gamma$-point.

Let us consider the emissivity dispersion diagrams near the $\Gamma$-point calculated for different photonic crystal periods, namely $a=525$\,nm, $a=600$\,nm and $a=675$\,nm (Figures\,\ref{spectra4}a--c). With an increase of $a$ the first Brillouin zone decreases and, as it can be seen from Figures\,\ref{spectra4}a--c, the spectral position of quasiguided modes redshifts. As a result, for different photonic crystal periods, different modes fall into the Ge nanoislands intrinsic emission range, bounded by the blue lines in Figures\,\ref{spectra4}a--c. This is in agreement with the experimental spectra shown in Figure\,\ref{spectra3} where some of the modes become invisible with an increase of $a$. Although there is a quite good match between the energy position of theoretically calculated modes and experimentally measured PL peaks (Figure\,\ref{spectra4}), some resonances manifest themselves in the measured PL spectra better than others. This is due to the fact that some of the resonances have almost a flat dispersion within 6\,$^\circ$ collecting cone, for example, the upper E$_1$ mode, while others have stronger dispersion such as the B$_1$ mode (the explanation of the formalism of the modes designation will be given later).

\begin{figure*}[t!]
\centering
\includegraphics[width=1\linewidth]{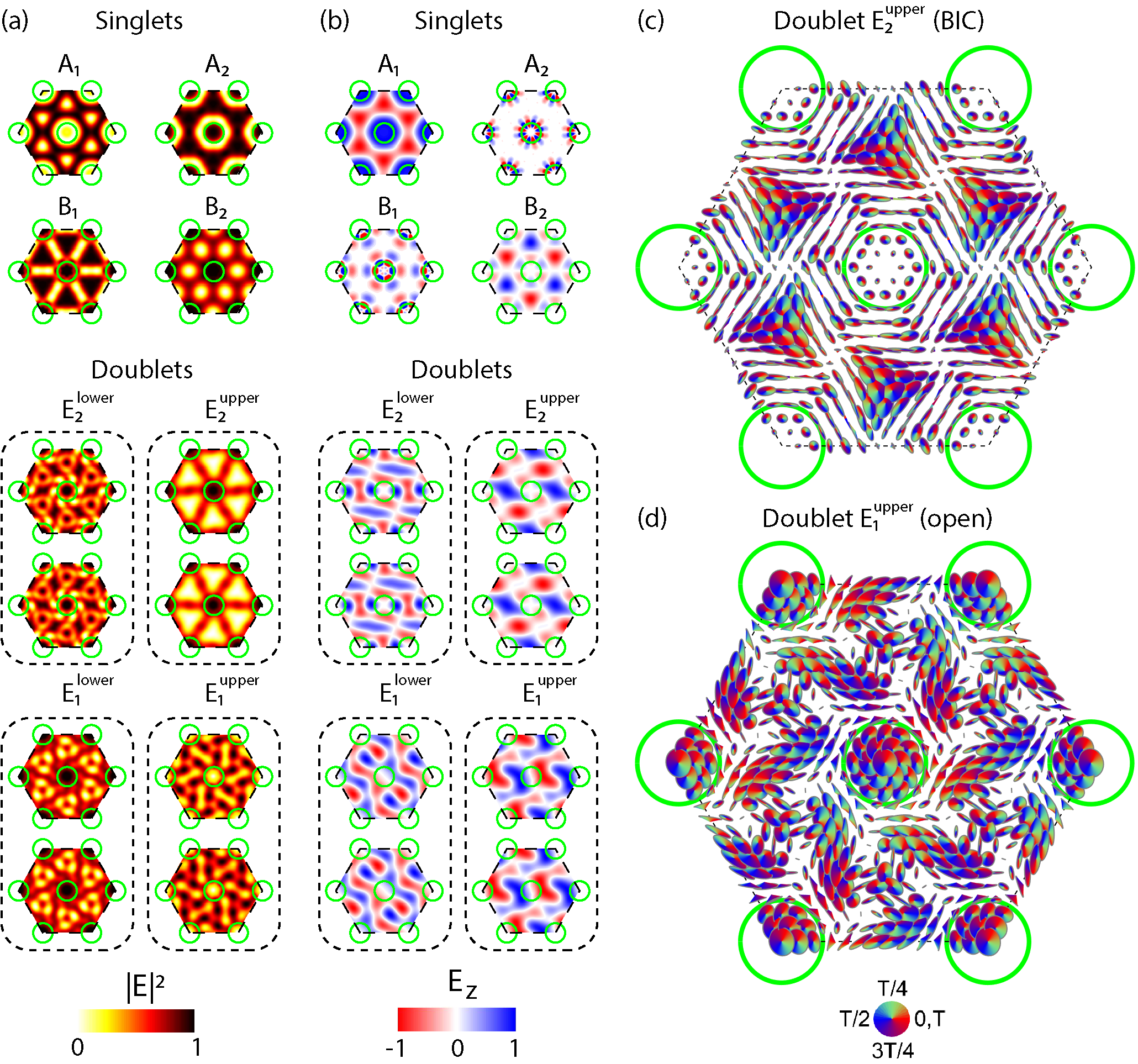}
\caption{(a) Electric field intensity and (b) $z$-projection of electric vector in singlet and doublet eigenmodes of PCS with $a=$\,600\,nm, $r/a=0.2$ calculated at $z$-coordinate in the middle of emitting layer. Photon energies of eigenmodes are the following: $\hbar\omega_{A_1} = 1088.3-0.705$i\,meV, $\hbar\omega_{A_2} = 843.9-0.439$i\,meV, $\hbar\omega_{B_1} = 812-0.458$i\,meV, $\hbar\omega_{B_2} = 898.4-0.609$i\,meV, $\hbar\omega_{E_1}^{\mathrm{upper}} = 960.3-2.770$i\,meV, $\hbar\omega_{E_1}^{\mathrm{lower}} = 916.2-5.263$i\,meV,
$\hbar\omega_{E_2}^{\mathrm{upper}} = 906.9-0.6$i\,meV,
$\hbar\omega_{E_2}^{\mathrm{lower}} = 837.6-0.434$i\,meV. Colorscales for panels (a--b) are shown on the bottom. Phase representation of electric field (c) in the upper doublet E$_2$ and (d) in the upper doublet E$_1$ for the PCS with $a=600$\,nm and $r/a=0.2$. The phase of electromagnetic oscillations is represented by color as shown on the circular color chart. Please note that (i) the cone base lies in the polarization plane where the field oscillates; (ii) the cone height is equal to the product of the electric field amplitude and the circular polarization degree; (iii) the direction of the cone follows the right screw rule.}
\label{fields}
\end{figure*}

As it has been mentioned above, the PL spectra of different PCSs have resonant peaks of different widths. This can be seen from the detailed spectra for PCSs with $a=525$\,nm, $a=600$\,nm and $a=675$\,nm shown separately in Figures\,\ref{spectra4}d--f. The experimental Q-factor of the peaks estimated as $Q=\lambda/\Delta\lambda$ is between 150 and 2200. To explain the difference of Q-factor in different modes, the symmetry of the quasiguided modes has to be considered. 

\section{Symmetry of modes}
The spatial symmetry of the quasiguided modes plays an important role in their optical response. The symmetry is determined by the type of the photonic crystal lattice and can be classified in terms of the group theory. To study the symmetry of the modes in our PCS, we find the poles of the scattering matrix by solving the eigenvalue problem (see Section Methods for details.). In the case of doublets, the solution is represented by the basis of two eigenfunctions; any linear combination of these eigenfunctions is a solution of the eigenvalue problem too. 

The spatial distributions of intensity and $z$-component of the electric field found from the output eigenvectors $\ket{\mathbf{O}}_{res}$ are shown in Figures\,\ref{fields}a and \ref{fields}b for the $\Gamma$-point. One can see that the intensity distributions in singlets are C$_{6v}$ symmetrical following the symmetry of the PCS. In the case of doublets, the intensity profile in the eigenmodes cannot be C$_{6v}$ symmetrical, however, one can choose a basis of eigenfunctions such that the field intensities in them have a C$_6$ symmetry as shown in Figure\,\ref{fields}a. The field intensities of two eigenfunctions are mirror-symmetrical to each other by a vertical plane, that indicates the degeneracy of these modes. The $E_z$ distributions in the singlets and doublets have more complex symmetries. The group theory defines symmetry of an eigenmode by a set of characters $\chi$ which characterize the mode transformation for each symmetry operation in the point group (see Table\,\ref{tabchar} in Appendix). By inspecting Figure\,\ref{fields}b one can associate each of the field distributions $E_z$ with a set of characters $\chi$ and, by this, determine to which irreducible representation of the C$_{6v}$ point group they belong. In Figure\,\ref{spectra4} and in Figure\,\ref{fields}, all 12 first-order guasiguided modes in the $\Gamma$-point are marked in accordance with the definitions in Table\,\ref{tabchar}. Please note that the $E_z$ distributions in both eigenfunctions of doublets are identical (bottom parts of Figures\,\ref{fields}a and \ref{fields}b).

In addition to the intensity and $E_z$ distributions, we present the phase distribution of electric field in the doublets E$_1$ and E$_2$ (Figure\,\ref{fields}c and \ref{fields}d) where the phase of electromagnetic oscillations is denoted by the color. Please see the details of this representation in Ref.\,\citenum{dyakov2018magnetic} and in Supporting Information. In Figure\,\ref{spectra4}c one can simultaneously see that (i) the presented field distributions are indeed the E$_1$ and E$_2$ modes; (ii) the electric vectors are not linearly polarized; (iii) the field intensity has C$_{6}$ rotational symmetry; (iv) the field in the doublets E$_2$ and E$_1$ has nodes and antinodes in the centers of pores. The phase representation of the fields for the rest of the modes is shown in Supporting Information. Figures\,\ref{fields}c and \ref{fields}d show only one of two eigenfunctions of doublets E$_1$ and E$_2$. The second eigenfunction of each doublet is mirror-symmetrical to the first eigenfunction with respect to a vertical plane passing through the centers of two nearest pores.

Understanding the symmetry of modes has a direct practical implication. Namely, it enables us to predict which modes can couple to the far field in the $\Gamma$-point. The coupling to free space is possible when the overlap integral $\gamma$ is non-vanishing:
\begin{equation}
    \gamma = \iint_{cell}\left(\vec{E}_{\mathrm{fs}}^*\times \vec{H}_{\mathrm{mode}}+\vec{E}_{\mathrm{mode}}^*\times \vec{H}_{\mathrm{fs}}\right)dS,
\end{equation}
where $dS$ is the element of the unit cell, indices $fs$ and $mode$ denote free-space and modal electric and magnetic fields, $E$ and $H$. By analyzing the characters of irreducible representations of C$_{6v}$ point group (Table\,\ref{tabchar} in Appendix) one can conclude that in the C$_{6v}$ symmetrical hexagonal lattice only the doublet E$_1$ is open for the far-field coupling. Whereas all the singlets, as well as the doublet E$_2$ are closed. Such modes are referred to as symmetry-protected (SP) bound states in the continuum \cite{Hsu2016a, marinica2008bound}. 

\begin{figure*}[t!]
\centering
\includegraphics[width=1\linewidth]{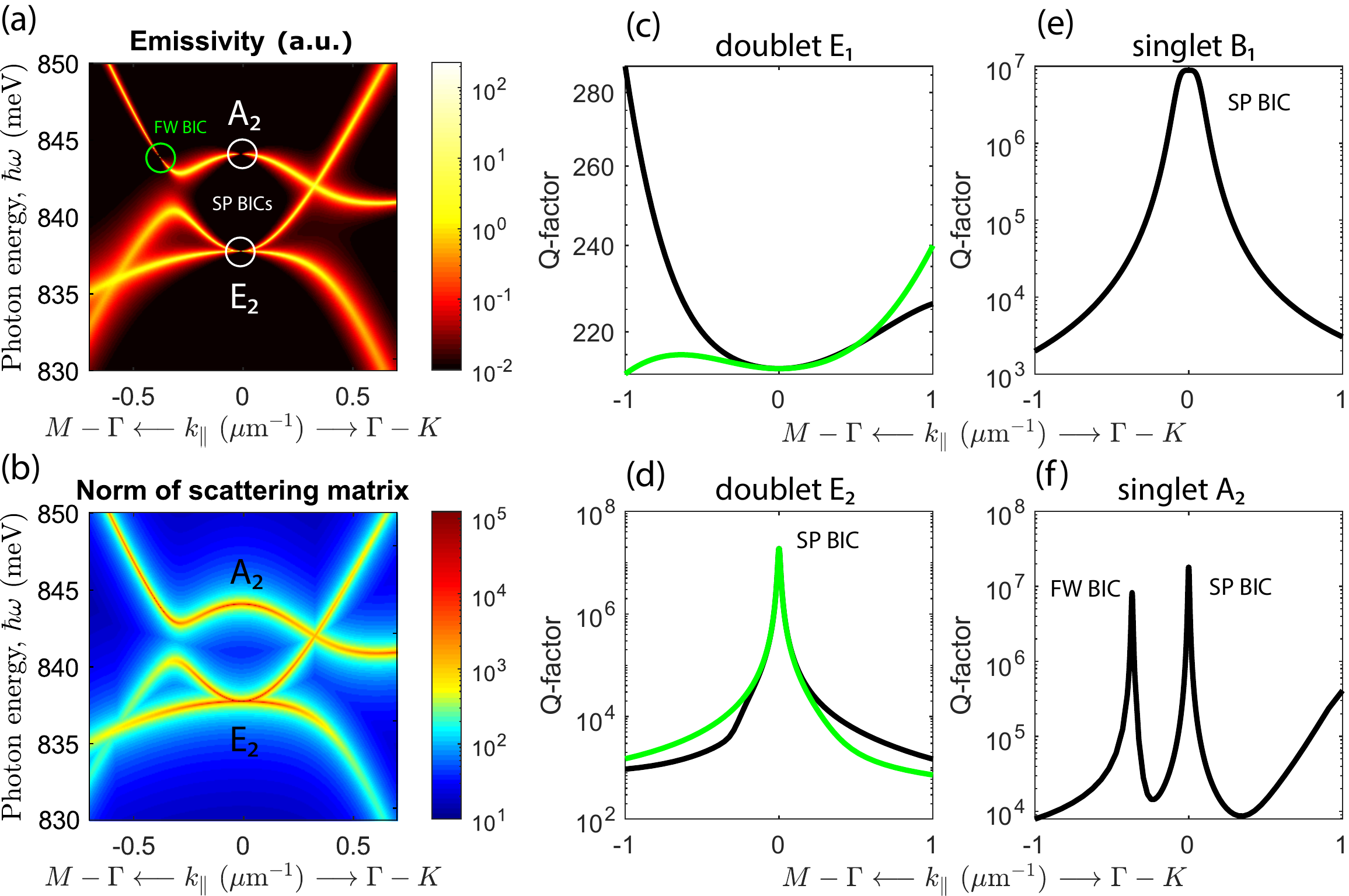}
\caption{(a) Calculated photon energy and wavevector dependence of (a) emissivity and (b) Frobenius norm of the scattering matrix near the $\Gamma$-point. Color scales are shown on the right. (c-f) Calculated Q-factor of selected resonances near $\Gamma$-point. Calculation for panels (a--f) are made for $a=$\,600\,nm and $r/a=$\,0.2 and effective refractive index of the layers with Ge nanoislands $n_{\mathrm{eff}}=3.12+10^{-6}i$. Symmetry-protected (SP) BIC and Friedrich-Wintgen (FW) BIC are highlighted.}
\label{norm}
\end{figure*}

\section{Bound states in the continuum}
As it has been mentioned, the group theory tells us that the number of symmetry protected BICs in a photonic crystal slab is fixed and is defined by the type of the photonic crystal lattice. In the C$_{6v}$ symmetrical hexagonal lattice, there are four singlet BICs and one doublet BIC irrespective of a particular geometry of the photonic crystal slab. By variation of geometrical parameters, one can only change the spectral position of modes in the $\Gamma$-point. As BICs cannot couple to the far field, they do not have radiation losses and, hence, the imaginary part of their frequency should be strictly zero in structures without Ohmic losses. In practice, we always have small but inevitable radiation losses caused by imperfections of geometry, roughness, non-periodicity, a limited number of periods in the photonic crystal slab, etc. It results in the fact that the Q-factor of such modes is not infinite, however, it can be very large. Yet another reason for the suppression of the experimental Q-factor of the BICs is that they might be measured out of the $\Gamma$-point due to experimental constraints. As we move away from the $\Gamma$-point, the symmetry of the BICs is broken and, consequently, they become visible in the far field. This is the reason why in our experiment we observe the peaks which can be associated with the BIC modes.

\begin{figure*}[t!]
\centering
\includegraphics[width=1\textwidth]{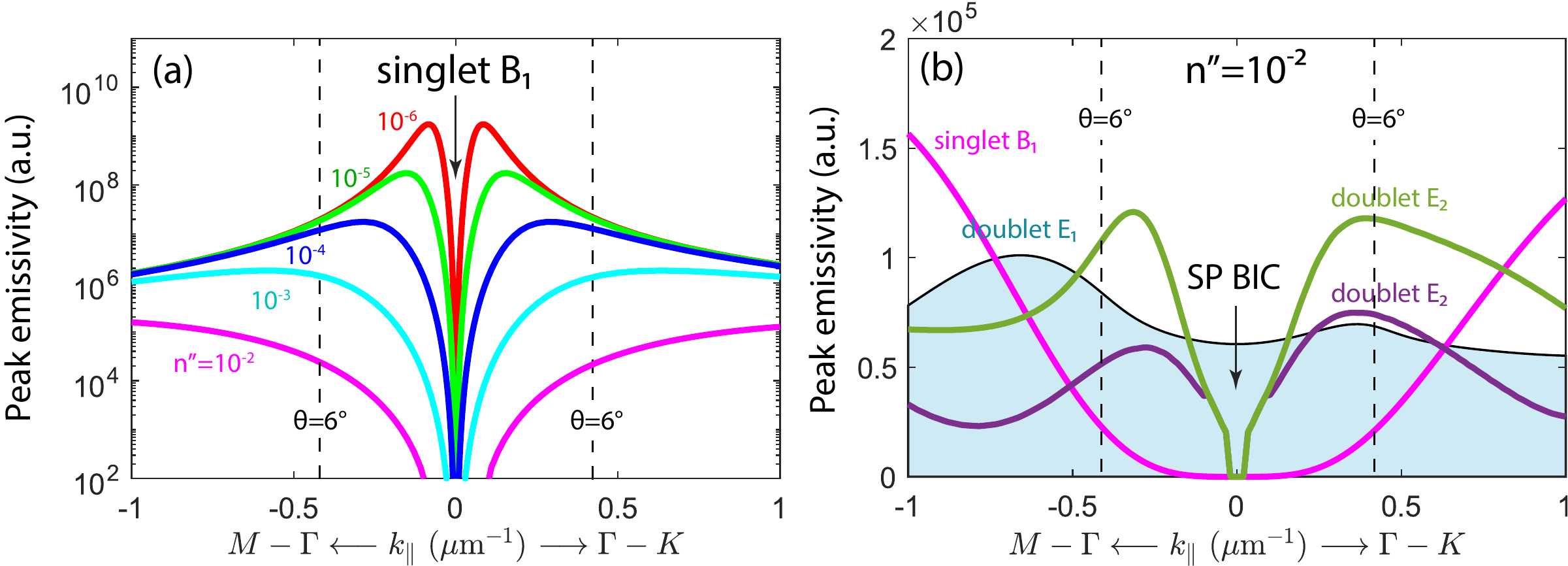}
\caption{In-plane wavevector dependence of the peak emissivity calculated near (a) the singlet B$_1$ for different imaginary parts of SiGe layer complex refractive index $n''$ and (b) near the singlet B$_1$ and doublets E$_1$ and E$_2$ at $n'' = 10^{-2}$. Black dashed lines in (a) and (b) denote the 6$^\circ$ light cone at $\hbar\omega=800$\,meV. The curves for the two modes near the doublet E$_1$ are very similar in the displayed range of $k_\parallel$, so that only one of them is shown. Real part of the complex refractive index is fixed at $n'=3.12$. The graphs in panels (a) and (b) should be interpreted in a manner that each point on a curve represents the photon energy where the emissivity reaches its maximum at certain value of $k_\parallel$. Note that the scale is logarithmic in (a) and linear in (b).}
\label{peakint}
\end{figure*}

To discuss this in more detail, we calculate the dispersions of the emissivity and the Frobenius norm of the scattering matrix in a narrow range of photon energy and in-plane wavevector near the modes A$_2$ and E$_2$ (Figures\,\ref{norm}a and \ref{norm}b). One can see that near the $\Gamma$-point the emissivity is suppressed as the modes A$_2$ and E$_2$ are the symmetry-protected BICs. Unlike the emissivity, the norm of the scattering matrix does not have similar discontinuities. Three local maxima at each $\vec{k}_\parallel$  represent the resonant poles; two of them degenerate in the $\Gamma$-point. The Q-factor of both of them in the $\Gamma$-point is extremely large as shown in Figures\,\ref{norm}d and \ref{norm}f. With an increase of $|k_\parallel|$ the degeneracy is lifted, and the Q-factor of A$_2$ and E$_2$ modes decreases. A similar in-plane wavevector dependence of the Q-factor can be observed for other BICs, as shown in Figure\,\ref{norm}e for the singlet B$_1$ as an example. The Q-factor of the open doublet E$_1$ in the $\Gamma$-point is sufficiently smaller than that of the BICs (Figure\,\ref{norm}c). This situation is experimentally demonstrated in Figures\,\ref{spectra3} and \ref{spectra4} where the peaks have different Q-factors. 

One can see in Figures\,\ref{norm}a and \ref{norm}b that there is yet another point in $k$-space, besides $\Gamma$, where the emissivity is suppressed and the resonance has exactly vanishing width, namely $k_\parallel=$\,0.4\,$\mu$m$^{-1}$ along the $\Gamma$-M direction. This is a BIC of Friedrich-Wintgen type \cite{friedrich1985interfering} which is a result of destructive interference between the modes with similar radiation patterns in far field \cite{gladyshev2018high}. When two quasiguided modes pass each other as a function of the in-plane wavevector, the interference causes an avoided crossing of the resonances and for a given value of $k_\parallel$ one resonance has extremely large Q-factor and, hence, becomes a BIC.

As the emissivity of BICs in the $\Gamma$-point is strictly zero, it appears that the possibility to obtain the BIC-original peaks in the PL spectra depends on the Ohmic losses power and on the solid angle from where the PL signal is collected. To demonstrate this we calculate the $k_\parallel$-dependence of the peak emissivity near the singlet $B_1$ for different parameters $n''$ which models the Ohmic losses \cite{dyakov2011optical} and assumes some fixed density of Ge nanoislands (Figure\,\ref{peakint}a). The parameter $n''$ is the imaginary part of the effective refractive index of the layer with Ge nanoislands. This parameter has definitely to be proportional to the Ge nanoislands density, but it also grows with the degree of overall disorder, introduced into Ge layers. The estimate for our samples is $n''=0.01$ which we have used in calculations of Figures\,\ref{EKX}--\ref{fields}. One can see that for all  $n''$ the peak emissivity rises with $|k_\parallel|$ reaching its maximum. With the increase of $n''$ the maximal peak emissivity moves away from the $\Gamma$-point. As a result, at $n''=0.01$, the total emissivity within the 6$^\circ$ light cone near the singlet B$_1$ is smaller than that near the dark doublet E$_2$ or bright doublet $E_1$, for instance (Figure\,\ref{peakint}b). It explains why the $B_1$ originated peak is poorly seen in the experimental DPL spectra unlike the other peaks (see Figure\,\ref{spectra4}). It is remarkable that as long as the absorption losses are small ($n''\leqslant 0.01$), the maximal peak emissivity is roughly proportional to $n''$. Hence, even at relatively high absorption (e.g., $n''=10^{-2}$), the peak PL becomes measurable provided that the PL signal is collected away of the $\Gamma$-point. Although, in the case of high absorption the quality factor of the resonance will be lower. Thus, in order to obtain high Q-factor resonance peaks one can use the advantage of BICs in lossless PCSs.


\section{Conclusion}
In conclusion, we have demonstrated experimentally that the photoluminescence of Ge nanoislands in silicon PCS with hexagonal lattice can be greatly enhanced by the symmetry-protected bounds states in the continuum. The peak enhancement factor reached 140 at the resonant frequencies and the overall integrated PL intensity was increased more by an order of magnitude as well.  In our PCS, the experimental PL peaks can be wide as well as narrow and exhibit sometimes a fine structure. We have theoretically simulated these spectral features by calculating the emissivity dispersion diagrams using the Fourier modal method in the scattering matrix form. On the dispersion diagrams, we have shown the appearance of singlet and doublet quasiguided modes in the $\Gamma$-point. We have classified these modes in terms of the group theory and have shown that in the C$_{6v}$ symmetrical PCS, the doublet modes can be optically dark (E$_2$) and bright (E$_1$). We have associated the experimentally observed peaks to certain irreducible representations of C$_{6v}$ point group. We also have shown that for the observation of BIC enhancement it is crucial to measure the photoluminescence response collecting the optical signal at angles close to normal. Finally, we have theoretically demonstrated the appearance of Friedrich-Wintgen BIC in our structure as a result of the destructive interference of two modes. 

\section{Methods}
\label{methods}
\subsection{Sample fabrication}

Photonic crystals were formed on structures with self-assembled Ge nanoislands grown by molecular beam epitaxy. An SOI wafer from SOITEC company with a 3\,$\mu$m thick oxide layer and a Si device layer thinned to 90\,nm was used as a substrate. Grown structures consist of a 75\,nm thick Si buffer layer, 5 layers of Ge nanoislands separated by 15\,nm thick Si spacer layers, and a 75\,nm thick Si cap layer. The total thickness of the structure above the buried oxide is 300\,nm. The islands were formed at 620$^\circ$C by the deposition of 7--8 monolayers of Ge. Under these growth conditions, an array of dome-shaped islands was formed on a silicon surface, with a surface density $\sim$10$^{10}$\,cm$^{-2}$, the lateral size of the islands of $\sim$70--80\,nm and the height before the overgrowth with the Si cap layer of $\sim$14--15\,nm. The choice of the growth temperature was determined by the results of earlier studies demonstrating that the structures with dome-shaped Ge islands formed at around 600$^\circ$C provide the highest emission intensity at room temperature \cite{vostokov2002low}. The luminescence properties of the “as grown” structures were analyzed in earlier works \cite{krasilnik2010sige}.

\begin{figure*}[t!]
\centering
\includegraphics[width=0.95\linewidth]{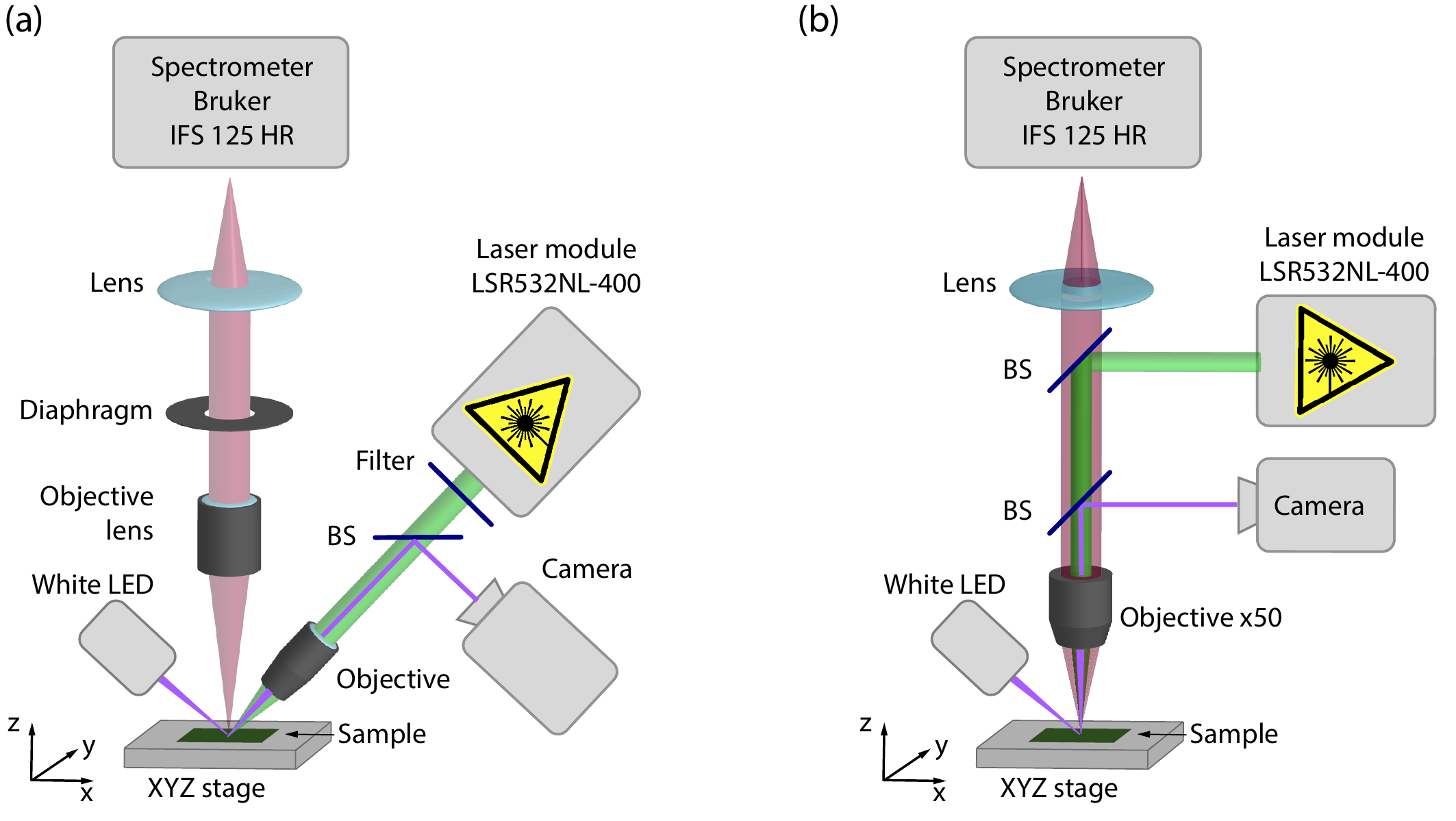}
\caption{Schematics of (a) the microphotoliminescence ($\mu$PL) setup and (b) directional photoluminescence (DPL) setup.}
\label{setup}
\end{figure*}

PCSs were formed by electron-beam lithography and plasma-chemical etching. At the first step, a PCS's pattern was formed in a PMMA resist using electron-beam lithography. This pattern served as a mask during the etching of the structure. Anisotropic etching of PCSs was performed using ICP plasma-chemical etching in SF$_6$/C$_4$F$_8$ mixture of gases. In this work, PCSs with a hexagonal hole lattice were studied. The PCS lattice period, $a$, was varied in the range from 450 to 725\,nm, and the ratio of the hole radius to the period, $r/a$, amounted to 0.2 and 0.26. The etching depth was 235\,nm for $r/a=0.2$ and 247\,nm for $r/a=0.26$. The overall size of the PCSs was 20$\times$25\,$\mu$m, thus PCSs contained more than 25$\times$25 periods.

\subsection{PL measurements setup}
We study the light-emitting properties of PCSs by applying two experimental techniques. 

With the first technique, one can measure the luminescence response of the photonic crystal structures with high spatial resolution. For this, we use a standard micro-photoluminescence ($\mu$PL) setup (Figure\,\ref{setup}a), where we collect the excitation light and the detecting signals through the same microscope objective, which provides better spatial resolution. For the microscope objective with $\times$50 magnification (Mitutoyo M Plan APO objective, NA=0.42), the spatial resolution amounts to approximately 2\,$\mu$m with the collection angles of up to $\sim$25$^\circ$ to normal.

To register the luminescence response at angles closer to the normal, we use a different setup, which we refer to as a directional photoluminescence (DPL) setup (Figure\,\ref{setup}b). The PL signal is excited by a laser beam with 60$^\circ$ incidence. A Mitutoyo M Plan APO objective with $\times$10 magnification focuses the laser beam on a spot with a diameter of $\sim$10\,$\mu$m. The PL signal is collected by the Nikon 50\,mm $f/1.4D$ AF Nikkor objective, which is located at the focal distance from the sample surface. In such a geometry, one can consider an emitted area as a point source. Accordingly, one can regard the light beam formed by the objective as parallel. The parallelism of the light beam enables us to use a diaphragm for collecting PL signals within small solid angles in the selected directions. We conduct PL measurements at the position of a diaphragm in the center of a parallel beam that corresponds to the maximal collection angle of $\sim$6$^\circ$ to normal.

We carry out all PL measurements at room temperature. The luminescence signal was excited by a solid-state CW laser emitting at the wavelength of 532\,nm (laser module LSR532NL-400). To detect the PL signal, we use a high-resolution Fourier spectrometer (Bruker IFS 125 HR) and a nitrogen-cooled Ge photodetector. The spectral resolution in both experimental schemes can reach 0.05 cm$^{-1}$.

\subsection{Theoretical method}
To theoretically study the optical behavior of the photonic crystal slab with Ge nanoislands, we use a Fourier modal method (FMM) in the scattering matrix form \cite{Tikhodeev2002b}, also known as rigorous coupled-wave analysis (RCWA) \cite{moharam1995formulation}. In the Fourier decomposition of electromagnetic fields, to preserve the C$_{6v}$ symmetry of the structure, we choose a C$_6$-symmetrical set of Fourier harmonics in the reciprocal space. The total number of harmonics is chosen to be $N_g = 199$ that ensures the convergence of our numerical scheme. As a result, we construct the $4N_g\times4 N_g$ dimensional scattering matrix $\mathbf{S}(\omega,\vec{k}_\parallel)$ which contains full optical information of our photonic crystal slab. Here $\omega$ and $\vec{k}_\parallel$ denote the frequency of electromagnetic oscillations and in-plane wavevector respectively.

We model the photoluminescence of Ge nanoislands by the radiation of chaotically oriented oscillating electric dipoles which is a good approximation of emitting molecules or quantum dots. To calculate the emissivity spectra of oscillating dipoles, we use the electrodynamic reciprocity principle. According to this principle the currents of two different dipoles $\vec{j}_{1,2}$ and their electric fields $\vec{E}_{1,2}$ at the positions of the other dipole are connected as $\vec{j}_1\vec{E}_2 = \vec{j}_2\vec{E}_1$. As a result, the problem of simulation of emissivity $I_i(\omega,\vec{k}_\parallel)$ of $i$-th dipole at the frequency $\omega$ and wavevector $\vec{k}_\parallel$ is reduced to the calculation of the electric near-field of a plane wave with the same $\omega$ and $\vec{k}_\parallel$ at the position of this dipole $\vec{r}_i$. The overall emissivity is found as a sum over the entire set of dipoles:
\begin{align}
    I(\omega,\vec{k}_\parallel) = 
    &\sum_i I_i(\omega,\vec{k}_\parallel) = \\ 
    &\sum_i|\vec{E}(\omega,\vec{k}_\parallel,\vec{r}_i)|^2,
\end{align}
where the subscript $i$ denotes the $i$-th dipole's current. We also average the emission over the polarization states distributed randomly between the following polarization vectors: 
\begin{align}
    \vec{p}_1 &= [1, 0]\\
    \vec{p}_2 &= [-1/2, -\sqrt{3}/2]\\
    \vec{p}_3 &= [-1/2, \sqrt{3}/2]
\end{align}
Such set of polarization vectors preserves the C$_{6v}$ rotational symmetry of the photonic crystal slab.

Please note that in this work we simulate the emissivity rather than a full photoluminescence response. The latter should include a non-homogeneous spatial excitation profile \cite{dyakov2018plasmon} which we omit here as we are focused on explaining the nature and symmetry of the resonances.

The eigenmodes of the photonic crystal slab are calculated by finding the poles of the scattering matrix \cite{Gippius2005c}. The corresponding eigenvalue problem can be written as
\begin{equation}
\mathbf{S}^{-1}(\omega,\vec{k}_\parallel)\ket{\mathbf{O}}_{res}=\ket{0},
\label{eigen}
\end{equation}
where, $\ket{\mathbf{O}}_{res}$ is the resonance output vector in the scattering matrix formalism (see Refs.\,[\citenum{Tikhodeev2002b, Gippius2005c}] for details). We solve the problem (\ref{eigen}) by the generalized Newton's method by means of linearization of the inverse scattering matrix in the complex frequency domain \cite{Gippius2005c}.

In FMM calculations, we describe the layer with Ge nanoislands by an effective refractive index $n_{\mathrm{eff}} = n' + n''i=$\,3.12+0.01i unless otherwise is stated. Dielectric permittivities of Si and SiO$_2$ are taken from Ref.\,[\citenum{palik1998handbook}]. In this work we use the convection exp($-\mathrm{i}\omega t$) for temporal dependencies of fields. In this convention photon energies of eigenmodes have negative imaginary parts.   

\subsection{Character table}
To describe the symmetry of structure eigenmodes, we use the notations from the group theory where the symmetry is defined as a set of characters. For singlets the characters are defined from
\begin{equation}
    \hat{R}E_z = \chi(\hat{R}) E_z,
\end{equation}
and for doublets
\begin{align}
    \hat{R}E_{1z} &= \chi_{11}E_{1z}+\chi_{12}E_{2z},\\
    \hat{R}E_{2z} &= \chi_{21}E_{1z}+\chi_{22}E_{2z},\\
    \chi(\hat{R}) &= \chi_{11} + \chi_{22},
\end{align}
where $\hat{R}$ denotes a symmetry operation in a point group. The table of characters for C$_{6v}$ point group is presented in Table\,\ref{tabchar}. 

\begin{table}[h]
\centering
\begin{tabular}{|l|r|r|r|r|r|r|} 
\hline
& E & 2C$_6$ & 2C$_3$ & C$_2$ & 3$\sigma_y$ & 3$\sigma_x$ \\ 
\hline
A$_1$  & 1 & 1 & 1 & 1 & 1 & 1 \\ 
\hline
A$_2$  & 1 & 1 & 1 & 1 & -1 & -1 \\ 
\hline
B$_1$  & 1 & -1 & 1 & -1 & 1 & -1 \\ 
\hline
B$_2$  & 1 & -1 & 1 & -1 & -1 & 1 \\ 
\hline
E$_1$  & 2 & 1 & -1 & -2 & 0 & 0 \\ 
\hline
E$_2$  & 2 & -1 & -1 & 2 & 0 & 0 \\ 
\hline
\end{tabular}
\caption{Character table for C$_{6v}$ point group.}
\label{tabchar}
\end{table}

\section{Acknowledgement}
This work was supported in part by the Russian Science Foundation (project 19-72-10011). The theoretical analysis of the modes emissivity in Sec. Bound states in the coontinuum was supported by the Russian Science Foundation (project \textnumero 16-12-10538$\Pi$). S.D. acknowledges I.M.\,Fradkin for fruitful discussions. A.B. acknowledges the BASIS foundation and Grant of the President of the Russian Federation (MK-2224.2020.2).


\end{document}